\documentclass[12pt]{iopart}
\usepackage{graphicx}
\usepackage{epsfig}
\bibstyle{apsrev.bib}

\begin{document}
\input epsf.sty

\title[Anomalous transport]{Anomalous transport in disordered exclusion processes with coupled particles}
\author{R\'obert Juh\'asz}
\address{Research Institute for Solid
State Physics and Optics, H-1525 Budapest, P.O. Box 49, Hungary}
\ead{juhasz@szfki.hu}

\begin{abstract}
We consider one-dimensional asymmetric exclusion processes with  
a simple attractive interaction, 
where the distance between consecutive particles is not allowed to exceed
a certain limit and 
investigate the consequences of this coupling 
on the transport properties in the presence of 
random-force type disorder by means of a phenomenological
random trap picture.
In the phase-separated steady state of the model defined on a finite ring,  
the properties of the density profile are studied  and
the exponent governing the decay of the current with the system size
in the biased phase is derived. 
In case all consecutive particles are coupled with each other and form a
closed string, the current is found to be enhanced compared to the model 
without coupling, while if groups of consecutive particles form finite strings, the current is reduced.    
The motion of a semi-infinite string entering an initially
empty lattice is also studied. Here, the diffusion of the head of the string
is found to be anomalous, and two phases can be distinguished, which are
characterised by different functional dependences of the
diffusion exponent on the bias.
The obtained results are checked by numerical simulation. 

\end{abstract}
\maketitle

\newcommand{\bc}{\begin{center}}
\newcommand{\ec}{\end{center}}
\newcommand{\be}{\begin{equation}}
\newcommand{\ee}{\end{equation}}
\newcommand{\beqn}{\begin{eqnarray}}
\newcommand{\eeqn}{\end{eqnarray}}

\vskip 2cm

\section{Introduction}
In low dimensions, the characteristics of transport in inhomogeneous
media may drastically differ from those in homogeneous environment.
This is to be seen already for the thoroughly-studied problem of random walk 
on a one-dimensional lattice with quenched random hop rates
\cite{alexander,havlin,haus,bouchaud,kesten1,derrida,sinai,fisher}.
The most striking anomalies are observed when the disorder is of 
random-force type, i.e. the direction of the local bias is random.  
In this case, the diffusion is anomalous, characterised by a diffusion
exponent continuously varying with the global bias
\cite{kesten1,derrida}, while it becomes
logarithmically slow for zero global bias---a phenomenon known as
Sinai diffusion \cite{sinai}. 
Much less is known for the transport of interacting many-particle
systems on disordered one-dimensional lattices.
The zero-range process (ZRP) \cite{spitzer,evanshanney}, 
where lattice sites are allowed to be multiply
occupied by identical particles, 
has a product-measure steady state even in the
case of random hop rates and it is closely related to the corresponding 
one-particle problem (random walk) on the same lattice. 
Beyond ZRP, perhaps the simplest interacting driven many-particle system is
the asymmetric simple exclusion process (ASEP)\cite{liggett,schutz}, 
where particles interact by hard-core exclusion.     
For this model, there are two ways to introduce disorder:
random rates can be associated either with particles or with sites
(links). 
The former model, the ASEP with particle-wise disorder
\cite{evans,letter,zero} can be mapped to a disordered ZRP, 
whereas for the ASEP with site-wise 
disorder \cite{stanley,tripathy,krug,jsi,barma}, general exact solutions are
not at our disposal. By means of a phenomenological random trap picture, 
the current of the latter model has been shown to display an 
anomalous behaviour similar to that of the random walk in the case 
of random-force type disorder \cite{jsi}. 

Hard-core exclusion can be thought of as a simple repulsive
interaction between neighbouring particles. One may also consider
attractive interactions between particles, 
the simplest form of which in one-dimension is,
analogous to hard-core exclusion, when the distance between
consecutive particles is not allowed to exceed a certain value, say, $l+1$ 
lattice spacings ($l\ge 1$), i.e. a string of coupled particles is formed.     
From the point of view of holes (empty sites),  
this prescription transforms to the constraint that the size of
clusters of holes is at most $l$. 
Regarding the number of holes in front of particle $i$ as an
occupation number on the $i$th site of a virtual lattice, the process
maps to the so-called generalised exclusion process \cite{liggett}, 
which is a ZRP with an upper limit $l$ for the local occupation numbers.  
Considering {\it particle-wise} disorder in the model of coupled particles, it
transforms to a generalised exclusion process with {\it site-wise} disorder.  
For the latter model with random-force type disorder, 
the scaling form of the current was argued
to be independent of $l$, thus identical to that of the ASEP, which
is the $l=1$ limit of the generalised exclusion process \cite{jsi}.  
The model of coupled particles with site-wise disorder 
with $l=1$ maps to an
exclusion process with double-sized particles with site-wise disorder,
however, a similar mapping does not exist for $l>1$ \cite{jsi}. 
For the process with $l=1$ in the presence of random-force type
disorder, the dynamical exponent has been found to differ from 
that of the ASEP \cite{jsi}. 

The aim of this paper is to study the exclusion process with coupled
particles in the presence of site-wise random-force type disorder 
for general $l$ and to explore, in which extent the transport
properties are modified compared to the ASEP by the simple attractive
interaction introduced above, which breaks particle-hole symmetry.   
We shall investigate the distribution of the sample-dependent current and the
properties of the density profile in the non-equilibrium steady state
on a finite ring, as well as  non-stationary phenomena such 
as the diffusion of a semi-infinite string by means of a 
phenomenological random trap model and numerical simulation.   

Beside theoretical interest, another source of motivations is the
applicability of simple variants of exclusion processes 
to the description of transport phenomena in a large variety of 
real systems such as vehicular traffic \cite{santen} or various 
biological transport processes \cite{chowdhury}.
Concerning the latter, random-force type disorder
may emerge in several contexts such as DNA unzipping \cite{lubensky},
translocation of RNA or DNA through pores \cite{kafri} or motion of molecular
motors \cite{howard} on heterogeneous tracks \cite{lipowsky,kafri}. 
In many cases, molecular motors act in groups and the collective
effects have been studied in different models \cite{julicher},
including models with hard core repulsion \cite{derenyi} or elastic coupling
\cite{csahok} between particles moving in a periodic potential.      
Our simple model may serve as a ground for testing the joint effect of
hard-core repulsion, coupling and disorder in such model systems.  
A more direct example is served by a recent work, 
in which the diffusion of finite strings of coupled particles on
a homogeneous one-dimensional lattice has been studied \cite{antal}, 
motivated by the modelling of synthetic molecular systems, known as
molecular spiders, which can move on surfaces and tracks \cite{pei}.  
The model to be studied here may thus be relevant for investigating the
influence of track heterogeneity on the transport properties of these systems.

The rest of the paper is organised as follows. In Sec. \ref{model}, the
model to be studied is defined. In Sec. \ref{steady}, 
the properties of closed strings
are investigated on finite rings and the results are formulated 
generally in Sec. \ref{general}. In Sec. \ref{diffusion}, 
the problem of diffusion of a semi-infinite string is discussed, while  
Sec. \ref{finite} is devoted to problems concerning the traffic of finite
strings, including the steady state and the invasion. 
Finally, the paper is closed with a discussion of the results 
in Sec. \ref{discussion}.

\section{The model}
\label{model}

The process that we focus on in the first part of this work is
defined on a ring with $L$ sites.  On this lattice, $N$ particles, 
which are numbered consecutively along the chain from left to right, are
distributed in such a way that each lattice site is occupied by at most
one particle and, in case of a closed string, for the number of empty
sites $h_i$ in front of the $i$th particle, $h_i\le l$ holds for all $i$,
while in case of an open string $h_i\le l$ is prescribed for all but one
particle (the $N$th one). 
Obviously, for the closed string, $L/(l+1)\le N$ must hold. 
In this system, a continuous-time Markov process is considered, in
the course of which particles attempt to change their positions 
independently and randomly. 
The allowed transitions are the following: The $i$th
particle on site $j$ attempts to hop to the adjacent lattice site on its
right-hand side with a site-dependent rate $p_j$, and
the trial is successful if $h_{i}\ge 1$ and $h_{i-1}<l$. Here,
$h_{0}\equiv h_N$ and for particle $1$ of an open string, the
condition $h_N<l$ is cancelled.
Particle $i$ on site $j$ hops to the adjacent site on its
left-hand side with a site-dependent rate $q_j$,      
provided $h_{i-1}\ge 1$ and $h_i<l$. For the $N$th particle of an open
string, the condition $h_N<l$ is ignored again. 
The hop rates $p_j$ and $q_j$ are independent, identically distributed
quenched random variables drawn from the distributions $\rho(p)$ and $\pi(q)$,
respectively. 
Introducing the potential difference between site $i$ and $i-1$ 
through the relation 
\be 
\Delta U_i\equiv \ln (q_{i}/p_{i-1}),
\label{pot}
\ee 
the average decrease of the potential per lattice spacing 
$F\equiv-\overline{\ln (q/p)}$ can be regarded as an average force
which the particles are subjected to. Here and in the rest of
the paper, the overbar stands for the average over $\rho(p)$ and
$\pi(q)$. Without loss of generality, we assume $F\ge 0$,
i.e. there is either an average bias to the right or the system is
unbiased and  
we restrict ourselves to randomness distributions where
the fraction of sites with $p_j<q_j$ is finite.   
The natural control parameter $\mu$ of the one-particle problem, which we
retain also for the many-particle process, is given by the
positive root of the equation (for $F>0$)\cite{bouchaud}:
\be 
\overline{\left(\frac{q}{p}\right)^{\mu}}=1.
\label{mugen}
\ee
It is monotonously increasing with $F$ and zero in the unbiased case ($F=0$).
In the numerical simulations, we used a bimodal distribution for the
hop rates,
where $p_iq_i=r$ holds for all $i$ and 
the distribution of forward hop rates $p_i$ is given by
\be 
\rho(p)=c\delta(p-r)+(1-c)\delta(p-1),
\label{bimodal}
\ee
with the parameters $0\le r\le 1$ and $0<c\le 1/2$. 
For this distribution, the average bias is zero if $c=1/2$ and the
control parameter reads as  
\be 
\mu = \frac{\ln (1/c-1)}{\ln (1/r)}.
\label{mu}
\ee 

\section{Steady state of a closed string}
\label{steady}

We start our investigations by analysing the stationary properties of
a finite closed string. 
In general, we are interested in the scaling behaviour
of various quantities in the large $L$ limit, when the global
density of particles $N/L$ is kept constant.

With the purpose of studying the model in 
the presence of an external bias ($\mu>0$), 
we invoke the random trap model of the
random walk in a biased random-force type environment. 
This picture is based on the observation that the walker spends 
long times in certain localised regions (trapping regions) and in
between, it performs a more or less directed motion. The process is thus
approximated by a directed walk between traps characterised by  
effective trapping (or waiting) times, the distribution of which is
broad \cite{bou89a}.
This simplified model proved to describe the large-scale properties of the
system correctly \cite{bouchaud}.
\begin{figure}[h]
\begin{center}
\includegraphics[width=0.6\linewidth]{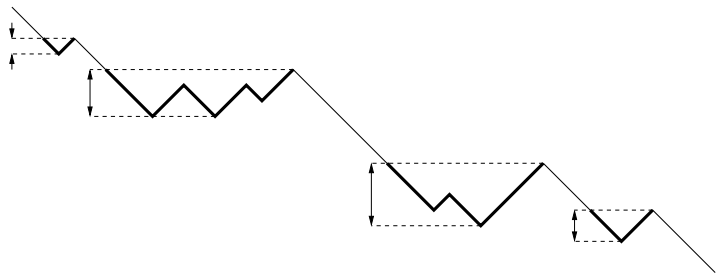}
\caption{\label{land} Illustration of the potential landscape defined
  by eq. (\ref{pot}). The trapping regions are indicated by thick lines
  and the arrows show the heights of the barriers.} 
\end{center}
\end{figure}
In the potential landscape of the infinite system, 
defined by eq. (\ref{pot}), which is a descending random walk path with average
slope $-F$ (see Fig. \ref{land}), a trapping region can be identified
as a basin on the left-hand side of a local maximum at site $n$ for
which $U_n>U_i$ holds for all sites $i>n$.  
The ascending section of the basin from the minimum to the maximum
will be termed barrier.  
The waiting time $\tau$ in a trapping region is in the order of the inverse
of the persistence probability of a walker starting at the minimum of the basin
with an imaginary absorbing site on its left-hand side \cite{jsi}. 
This quantity is of the form of a Kesten-variable \cite{kesten2}, the
distribution of which has an algebraic tail
for large $\tau$:
\be 
p(\tau)\sim\tau^{-1-\mu},
\label{taudist}
\ee
where $\mu$ is given by eq. (\ref{mugen}). 
The waiting time $\tau$ is related to the potential difference between
the maximum and minimum, i.e. the height of the barrier $U$ 
as $\tau\sim e^{U}$ for large $U$.
For the ASEP with site-wise disorder, which has been studied in the framework of
this simplified model \cite{jsi}, a phase separation can be observed: 
At the largest barrier, which serves as a bottleneck, 
a front appears, which  
separates a macroscopic high-density domain from a macroscopic low-density
one and, as a consequence of
particle-hole symmetry, the front is located where the potential is
half the height of the barrier \cite{blythe}.
For the process under study, the situation is similar: the current is
controlled by the largest barrier present in the system, where a front
develops, however, as the particle-hole symmetry is broken,
half-filling is no longer valid.
Therefore, we consider first an isolated barrier of height $U$ and express the
current in terms of the one-particle waiting time.     

\begin{figure}[h]
\begin{center}
\includegraphics[width=0.5\linewidth]{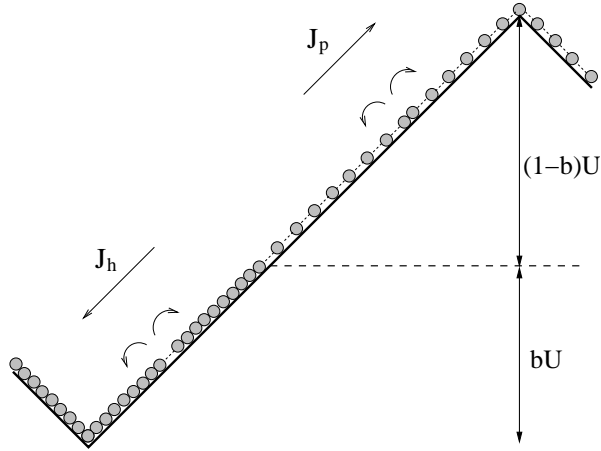}
\caption{\label{barrier} Typical configuration of particles at an
  isolated barrier for $l=1$.} 
\end{center}
\end{figure}

Let us assume that the front is located at some potential $bU$ $(b<1)$, 
which is measured from the
bottom of the barrier (see Fig. \ref{barrier}). On the left-hand side of the front
the particles sit closely next to each other so that the 
particle density is nearly $1$ (high-density phase), whereas on the right-hand side, particles are as far
from each other as possible and the density is close to $1/(l+1)$
(low-density phase). 
In case of such a configuration, two kinds of processes which result in
the shift of the front may take place: 
Starting from the front, holes diffuse to the left through
the high-density phase and particles diffuse from the front to the right
through the low-density phase over the barrier.  
In fact, the event that a hole detaches from the front and goes
away from it is very rare: the probability for this decreases exponentially
with the potential difference. 
Therefore the diffusing holes are
practically alone in the high-density phase far from the front and   
the typical time scale for a hole to successfully 
overcome the potential barrier $bU$ and escape to the left is in the 
order of $e^{bU}$. 
The current of holes in the high-density region is thus 
$J_h\sim e^{-bU}$.  
The mechanism of particle transport in the low-density phase is
similar. Here, particles are immobile since the interparticle spacings 
are almost everywhere maximal ($h_i=l$). 
If the particle at the front jumps to the right, 
a short interparticle spacing with $h_i=l-1$ arises in front of
this particle and  
thereby makes it possible for the next particle on its right-hand side, to
jump to the right. If the latter event occurs, it can be regarded as if
the short spacing moved to the
right. For similar reasons as for the holes in the high-density phase,
a short spacing is typically by oneself in the low-density
phase and  moves like a random walker taking steps of length of
$l+1$ lattice spacings. Thus, in order to escape to the right, this walker
has to overcome only a reduced potential barrier
$(1-b)U/(l+1)$. The corresponding time scale is
$e^{(1-b)U/(l+1)}$ and the current of particles through the
low-density phase is $J_p \sim \frac{1}{l+1}e^{-(1-b)U/(l+1)}$. 
In the steady state, $J_h=J_p$ must hold, otherwise the front would
move with a finite velocity. This leads to 
$e^{-bU}\sim \frac{1}{l+1}e^{-(1-b)U/(l+1)}$ and for large $U$, 
i.e. $U\gg l\ln l$, we obtain $b=1/(l+2)$. 
The particle current $j$ at the barrier is thus given by 
\be 
j\sim e^{-U/(l+2)} \sim \tau^{-1/(l+2)}
\label{j}
\ee 
in terms of the one-particle waiting time $\tau$.

\subsection{Current}

It is clear that the quantity $j$ derived above is merely 
the transport capacity
of the barrier, i.e. the maximal current   
which can flow through it and the actual current is
controlled by the incoming current of particles $j_{in}$ if $j_{in}<j$.  
In a disordered ring of size $L$, the stationary current $J_L$ is 
determined by the smallest one among the capacities of barriers present in the
system: $J_L=\min\{j_i\}$.  
Making use of eq. (\ref{taudist}) and eq. (\ref{j}), 
we obtain that the distribution of capacities has a power-law tail 
$\rho(j)\sim j^{-1+(l+2)\mu}$ for small $j$. The distribution of the
sample-dependent current $J_L$ is thus given by
the well-known Fr\'echet distribution for large $L$ \cite{galambos}:
\be      
p(\tilde J)=(l+2)\mu \tilde J^{(l+2)\mu-1}e^{-\tilde J^{(l+2)\mu}},
\label{frechet}
\ee
in terms of the scaling variable $\tilde J=cJ_LL^{\frac{1}{(l+2)\mu}}$, 
where the constant $c$ is
related to the pre-factor in the asymptotical form of $\rho(j)$. 
Thus, the current scales with the system size $L$ as 
\be 
J_L\sim L^{-\frac{1}{(l+2)\mu}}, 
\label{Jscale}
\ee
and vanishes in the limit $L\to \infty$. 

\begin{figure}[h]
\begin{center}
\includegraphics[width=0.9\linewidth]{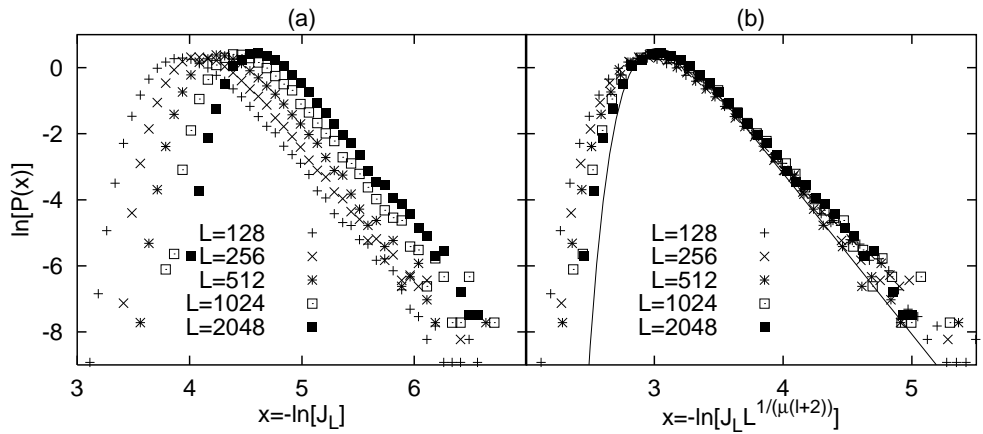}
\caption{\label{c1} a) Distribution of the logarithm of the current 
measured in numerical 
simulations for different system sizes in the model with $l=2$. 
b) Scaling plot of the
distributions. The number of particles was $N=L/2$ and the 
binary randomness defined in eq. (\ref{bimodal})
was used with $c=0.3$ and $r=0.5$, where the control parameter is
$\mu \approx 1.222$. The solid curve is the Fr\'echet distribution
given in eq. (\ref{frechet}).} 
\end{center}
\end{figure}

\begin{figure}[h]
\begin{center}
\includegraphics[width=0.9\linewidth]{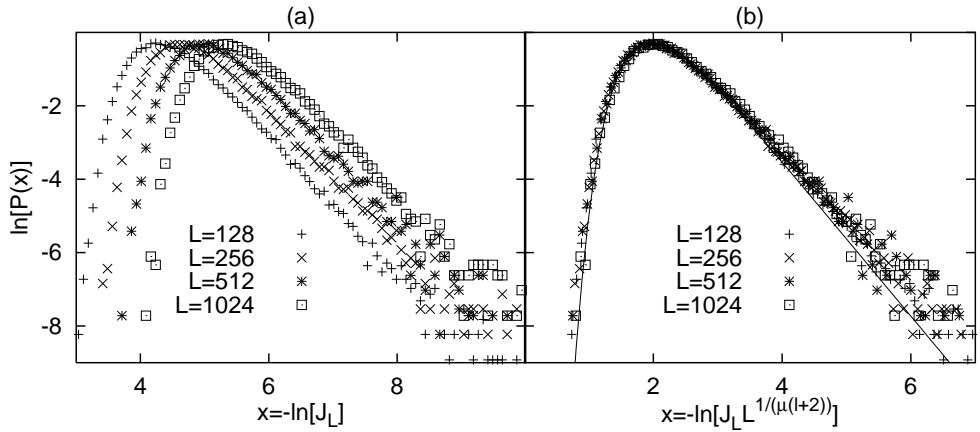}
\caption{\label{c2} 
a) Distribution of the logarithm of the current 
measured in numerical 
simulations for different system sizes in the model with $l=2$. 
b) Scaling plot of the
distributions. The number of particles was $N=L/2$ and the 
binary randomness defined in eq. (\ref{bimodal})
was used with $c=0.3$ and $r=0.2$, where the control parameter is
$\mu \approx 0.526$. The solid curve is the Fr\'echet distribution
given in eq. (\ref{frechet}).} 
\end{center}
\end{figure}

We have performed numerical simulations for finite rings of size
$L=128,256,512,1024,2048$ and after waiting 
sufficiently long time such that the
system has settled in a steady state we measured the 
current. This procedure was then repeated for $3\times 10^4$ 
independent samples for each $L$. As can be seen in
Fig. \ref{c1} and Fig. \ref{c2}, 
the Fr\'echet distribution fits satisfactorily to 
the distribution of the current, and 
keeping in mind that the size of the largest trapping region 
is only $O(\ln L)$, the deviations
can be assigned to corrections to scaling, which may be still
considerable for the numerically available systems sizes.      

Reducing the control parameter, the exponent governing the finite-size
scaling of
the current is increasing and finally diverges 
as the unbiased situation is approached, i.e. when $\mu\to 0$.  
Strictly for $\mu =0$, the random trap approximation breaks down since the
size of the largest trapping region is $O(L)$. 
Since the height of the largest barrier is $O(\sqrt{L})$, we
expect that the magnitude of the current scales typically as 
$-\ln|J_L|\sim \sqrt{L}$. 
As the relaxation time is exponentially large,
the dynamics in the unbiased case will be tested numerically in the
context of the diffusion of a semi-infinite string in Sec. \ref{diffusion}.  

\subsection{Density profile}
 
\begin{figure}[h]
\begin{center}
\includegraphics[width=0.8\linewidth]{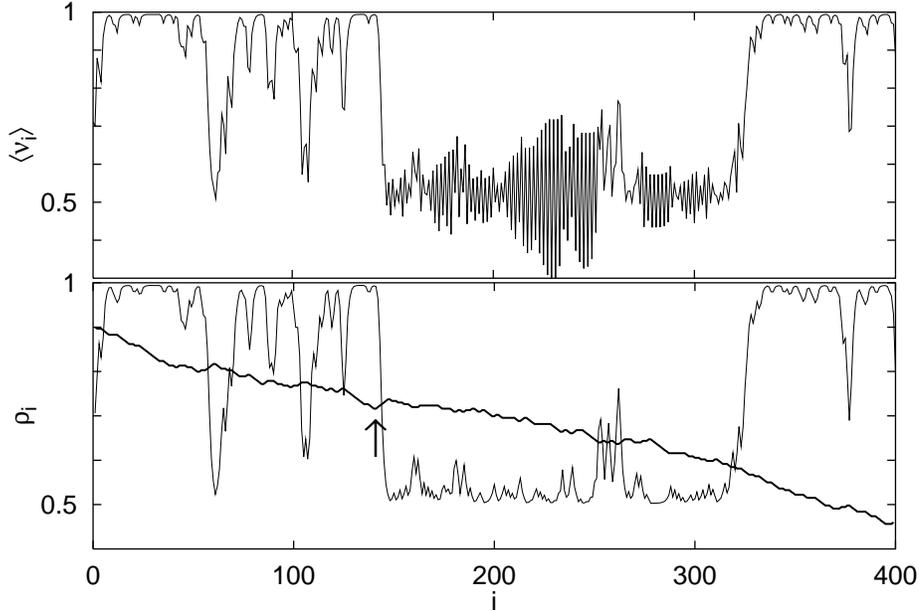}
\caption{\label{sample} Upper panel: Steady-state density profile 
of the model with $l=1$ obtained by numerical simulation in a sample
 of size $L=400$. 
The number of particles is $N=300$ and the rates were drawn from the
 bimodal distribution with $c=1/3$ and $r=1/4$. Lower panel: Smoothed
profile, obtained by averaging over pairs of sites: $\rho_i=(\langle
 \nu_i\rangle+\langle \nu_{i+1}\rangle)/2$. The thick line is the
 potential landscape defined by eq. (\ref{pot}) and the arrow
 indicates the starting point of the largest barrier.} 
\end{center}
\end{figure}

The steady-state average of the local
occupation number $\nu_i$, which is zero (one) for empty (occupied) sites, 
is plotted against the site index $i$ in Fig. \ref{sample} 
for a given sample.   
As can be seen, the steady state is segregated: the profile
consists of a low-density phase, which extends from site $\sim 150$
to site $\sim 320$, and a high-density phase in the remaining part
of the ring. In the low-density phase, the string is stretched out
(i.e. $h_i=l$) almost everywhere and this configuration is more or less 
fixed by the rugged landscape, which results in strong spatial variations
in the profile with a period $l+1$. After filtering out
this noise (Fig. \ref{sample}, lower panel), peaks can be observed in
the profile, where the local density significantly exceeds the
background density $1/(l+1)$. Similarly, in the high-density phase gaps
with low local density develop. 
In the following, we shall estimate the number of peaks and gaps and
will see that they both grow algebraically with $L$, however, with different
exponents. 

The peaks and gaps are accumulations of particles and holes,
respectively, at barriers with sufficiently large one-particle
waiting times. 
First, let us consider a barrier in the low-density phase with a trapping time
$\tau$ and assume that the string is completely stretched out here,
i.e. the interparticle spacings are maximal. Recalling the
considerations of the previous section, the capacity of
the barrier in case of this configuration is in the order of 
$\frac{1}{l+1}\tau^{-1/(l+1)}$. Apparently, this configuration remains 
stable only if the incoming flow of particles $J_L$, which is
determined by the largest barrier, does not exceed the capacity of the
barrier concerning this configuration. In the opposite case, particles
gradually accumulate here and form a high-density cluster. 
This process goes on until the front rises
up to a certain potential level, so that the capacity concerning this new
configuration is identical to the current $J_L$. (In this case, the
magnitude of the potential at the front measured from the top of the
barrier is $\frac{l+1}{l+2}U$, where $U\sim (l+2)\ln(1/J_L)$ 
is the height of the largest barrier.)        
We thus expect an accumulation of particles at those trapping regions
where $\frac{1}{l+1}\tau^{-1/(l+1)}<J_L$ holds, or, $\tau >
J_L^{-(l+1)}$ for large $L$. 
The number of barriers of this property is 
$N_{l}\sim L\int_{J_L^{-(l+1)}}^{\infty}p(\tau)d\tau \sim
LJ_L^{(l+1)\mu}$, where we made use of eq. (\ref{taudist}).    
Using eq. (\ref{Jscale}), we obtain for the number of peaks in
the low-density phase:
\be 
N_l\sim L^{\frac{1}{l+2}}.
\label{nl}
\ee

\begin{figure}[h]
\begin{center}
\includegraphics[width=0.8\linewidth]{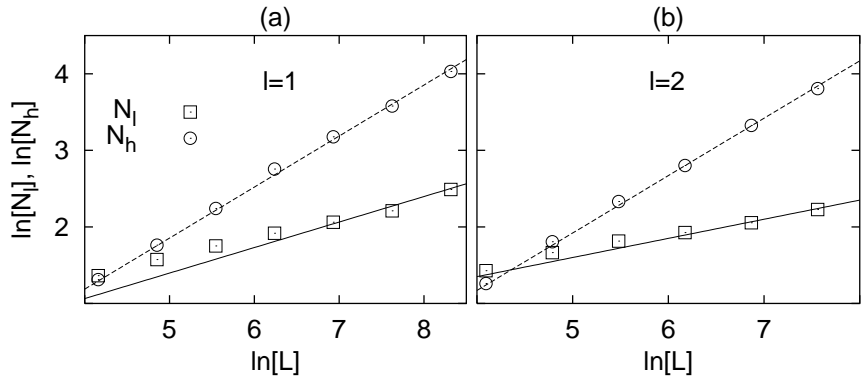}
\caption{\label{cross} Size dependence of the average number of intersections
  (defined in the text) obtained by numerical simulation for (a) $l=1$
  and (b) $l=2$. Bimodal randomness was used with
  $c=1/3$ and $r=1/4$ and the averaging was performed over $10^3$
  samples. The slope of the solid and dashed lines is $\frac{1}{l+2}$
  and $\frac{l+1}{l+2}$, respectively.} 
\end{center}
\end{figure}

In the high-density phase, the transport is realized by the diffusion 
of vacancies, and an accumulation, i.e. formation of clusters 
of $l$ holes separated by a particle occurs at barriers for which 
$\tau > J_L^{-1}$ holds. By a calculation analogous to that carried out for
the peaks, we obtain that the number of gaps in the high-density phase is 
\be 
N_h\sim L^{\frac{l+1}{l+2}}.
\label{nh}
\ee 
As can be seen, the exponents in eq. (\ref{nl}) and eq. (\ref{nh}) depend
only on the parameter $l$ and are independent of $\mu$.
In order to check these findings, we have performed numerical
simulations with a global density $N/L=\frac{l+2}{2(l+1)}$, so that
both the length of the low-density phase and of the high-density phase was
$O(L/2)$. 
It was measured in both phases how many times the smoothed steady-state
profile intersects the horizontal line at the density $\frac{l+2}{2(l+1)}$.
The results shown in Fig. \ref{cross} are in satisfactory agreement
with the theoretical predictions.    

\subsection{Active particles, active holes}
\label{ap}

In a given time in the steady state, some of the
particles reside in the high-density phase and
some of them reside in one of the particle clusters (peaks) in the
low-density phase. Almost all of these particles are blocked by the
hard-core exclusion. Other particles are located in the low-density
phase and are blocked since the string is stretched out. 
All these particles may be termed inactive as they do not contribute
to the current. 
On the other hand, there is a vanishing fraction of particles 
located between peaks in the low-density phase, which are
not blocked by adjacent particles and which are responsible for
the transport. In fact, the
transport is realized by the diffusion of short interparticle
spacings, which we shall call {\it active particles}. 
Analogously, in the high-density phase, the current is ascribed to
vacancies which diffuse freely between gaps, the so called 
{\it active holes}. 
In the following, we shall estimate the number of active particles and
active holes present in the system.   

In the high-density phase, the typical distance between adjacent gaps
is $\xi_h\sim L/N_h\sim L^{\frac{1}{l+2}}$. We will see a posteriori
that the concentration of holes in these domains is vanishing, so that
they can be regarded as independent walkers. 
It is known that the motion of a single random walker is 
controlled by the parameter $\mu$ \cite{bouchaud}. 
If $\mu >1$, the velocity of active holes is constant, so they pass through
the inter-gap region in a time $t\sim \xi_h$. The number of active holes
$N^a_{h}$ present in this domain is related to the current as  
$N^a_{h}\sim J_Lt$. Using eq. (\ref{Jscale}), we obtain for the number of
active holes between adjacent gaps: $N^a_{h} \sim L^{\frac{1-1/\mu}{l+2}}$. 
If $\mu <1$, the diffusion of holes is anomalous, i.e. the travelling
time is  $t\sim \xi_h^{\frac{1}{\mu}}$
and  the number of active holes between adjacent gaps is independent
of $L$: $N^a_{h}\sim J_L\xi_h^{1/\mu}\sim O(1)$. 
The concentration of active holes $N^a_{h}/\xi_h$ is thus indeed 
vanishing for any $\mu$.    

In the low-density phase, the typical distance between neighbouring
peaks is $\xi_l\sim L/N_l\sim L^{\frac{l+1}{l+2}}$. 
The active particles (the short spacings) take steps of length $l+1$, 
therefore their effective waiting time at a barrier is $\tau^{1/(l+1)}$. 
In the asymptotical form of the distribution of this quantity, 
$(l+1)\mu$ appears in place of $\mu$,  
thus, we conclude that 
the motion of active particles is governed by the effective
control parameter $(l+1)\mu$. 
If $\mu > \frac{1}{l+1}$, the velocity of active particles is finite
and we obtain for their number between two adjacent peaks: 
$N^a_{p}\sim J_L\xi_l \sim L^{\frac{l+1}{l+2}-\frac{1}{(l+2)\mu}}$.   
If $\mu < \frac{1}{l+1}$, we have $t\sim \xi_l^{\frac{1}{(l+1)\mu}}$
and the number of active particles is independent of $L$:
$N^a_{p}\sim J_L\xi_l^{\frac{1}{(l+1)\mu}} \sim O(1)$. 
As can be seen also the concentration of active particles 
$N^a_{p}/\xi_l$ is vanishing.  

We have obtained that, as opposed to the number of peaks and gaps, 
the numbers of active holes and particles depend also
on $\mu$ and the behaviour of the former
changes at $\mu=1$ while the behaviour of the latter changes at $\mu =\frac{1}{l+1}$.     

\section{General formulation of steady-state results}
\label{general}

The results obtained in the previous section are easy to formulate
generally for exclusion processes with random-force type disorder, 
where the active particles and holes responsible for the
transport in the low-density and high-density phase (which are not necessarily
true particles or holes, as we have seen above) 
overcome barriers according to the Arrhenius
type activated dynamics characterised by a waiting time $\tilde \tau_i\sim e^{f_iU}$.  
Here, $i=1,2$ and the index $1$ ($2$) refers to active holes
(particles) in the high-density (low-density) phase, and the factors $f_i$
are characteristic of the particular system. 
This class of models includes the ASEP, the closed string, the
ASEP with large particles \cite{jsi}, the model with finite open
strings to be discussed in Sec. \ref{finite}, and  possible
combinations of these models. 

For models of this class, all the exponents characterising the stationary
behaviour of quantities
studied in the previous section can be expressed in terms of two
exponents: $\mu_1= \mu/f_1$ and $\mu_2= \mu/f_2$, which
appear in the distribution of effective waiting times $\tilde\tau_i$ 
of active particles and
holes: $\tilde p_i(\tilde\tau_i)\sim\tilde\tau_i^{-1-\mu_i}$, $i=1,2$. 
Generalising the argumentations presented in the previous section, 
one can easily show that the steady state current follows a
Fr\'echet distribution and vanishes
according to 
\be 
J_L\sim L^{-\frac{1}{\mu_1+\mu_2}}.
\label{Jgen}
\ee
The number of gaps in the high-density phase ($i=1$) and the number of
peaks in the low-density phase ($i=2$) scales with the system size as 
\be 
N_i\sim  L^{1-\frac{\mu_i}{\mu_1+\mu_2}}.
\label{ngen}     
\ee
The typical number of active particles and holes in an
inter-barrier domain is 
\be 
N^a_{i}\sim \left\{
\begin{array}{c}
L^{\frac{\mu_i-1}{\mu_1+\mu_2}} \qquad \mu_i>1 \\
O(1). \qquad \mu_i<1.
\end{array}
\right. 
\label{agen}
\ee
For the closed string, the basic exponents are $\mu_1=\mu$ and
$\mu_2=(l+1)\mu$; for the ASEP, $\mu_1=\mu_2=\mu$; and for the
exclusion process with particles of size $d$, which was studied in 
Ref. \cite{jsi}, $\mu_1=d\mu$ and $\mu_2=\mu$. 

\section{Diffusion of a semi-infinite string}
\label{diffusion}

\begin{figure}[h]
\begin{center}
\includegraphics[width=0.3\linewidth]{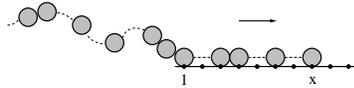}
\caption{\label{invasion} Advance of a semi-infinite string.} 
\end{center}
\end{figure}

So far, we have studied the steady state properties of the model defined 
on a ring. 
Now, we examine how a semi-infinite string moves when it enters an
initially empty semi-infinite lattice (see Fig. \ref{invasion}). 
To be concrete, the string enters at the first lattice site, 
which means formally that particles are created here with
rate $1$ and the particle closest to the entrance site is not allowed
to jump to the right if it resides at site $l+1$. 
We are interested in the time-dependence of the position 
$\overline{\langle x\rangle}$ of the first particle 
(the head) of the string.
Here, the angular brackets denote average over stochastic
histories.  
Let us consider first an isolated barrier of height $U$ and examine the way the
head of the string goes over it.
 
\begin{figure}[h]
\begin{center}
\includegraphics[width=0.5\linewidth]{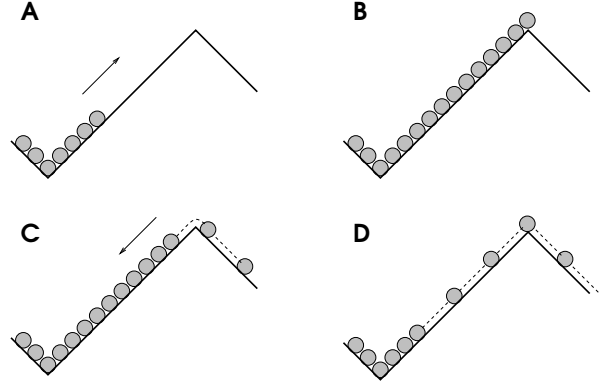}
\caption{\label{inv1} The mechanism of overcoming a barrier. } 
\end{center}
\end{figure}

\begin{figure}[h]
\begin{center}
\includegraphics[width=0.8\linewidth]{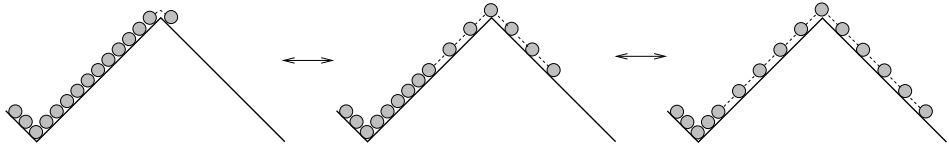}
\caption{\label{l1inv} The ``swinging'' of head of the string in the
  case $l=1$. } 
\end{center}
\end{figure}

First, the front of the high-density cluster of accumulated
particles is advancing (Fig. \ref{inv1}A) until the head
reaches the top of the barrier (Fig. \ref{inv1}B). This process is
realized by diffusion of vacancies through the high-density cluster
and its time scale is dominated by the
escape time of vacancies when the head is in the vicinity of the top, 
which is $O(e^U)$.    
After the top was reached the head falls over and a stretched segment
develops while the front separating it from  the high-density segment 
moves backwards (Fig. \ref{inv1}C). This
process, which is realized by the escape of active particles 
through the stretched segment, goes on until the stable level of 
the front determined by the incoming current is reached (Fig. \ref{inv1}D). 
Denoting the potential difference between this level and
the maximum by $U_0$, the time scale of this process is $O(e^{U_0/(l+1)})$. 
While for $l>1$, the motion of the head is more or less
directed after it has fallen over the maximum, for $l=1$ a special
phenomenon can be observed: The head performs an unbiased
random walk and may even return to the top (see Fig. \ref{l1inv}). 
As can be seen from the configurations shown in the figure, the
stretched segment is located symmetrically on the two sides of the
barrier hence the active particles go over the barrier in both
directions with the same rate. Thus, the motion of the head can
be described by a symmetric random walk with hop rates 
$p(U')=q(U')\sim e^{-U'/(l+1)}$, which depend on the magnitude of potential
$U'$ (measured from the maximum) at the location of the head.
The characteristic time in which the
head first reaches a site of potential $U'$ is in the
order of $e^{U'/(l+1)}$, see e.g. Ref. \cite{murthy}.     
However, during the swinging of the head, 
vacancies may also escape to the left through the high-density
segment. These processes result in the shift of the front and thus
induce a bias to the right for the motion of the head. The swinging
therefore goes on practically until the front first reaches its stable
level and afterwards the motion of the head becomes directed. 

We see that the time-determining one among the above steps is the
climb of the head to the top of the barrier, the characteristic time
of which is in the order of the one-particle waiting time $e^{U}$.
It is, however, valid only if the incoming current of
particles $j_{in}$ exceeds $\tau^{-1}$. 
Otherwise the time scale of overcoming the barrier 
is obviously $O(\xi/j_{in})$, where $\xi$
is the size of the trapping region.    

Now, we return to the investigation of the disordered model, where the
potential landscape contains barriers of various heights. 
In what follows, we number the barriers starting from the
first site of the lattice and measure distances in 
terms of the barrier index $n$. This can be done since the average distance 
between the starting points of consecutive barriers is finite. 
It is clear that $j_i(l+1)$ is an upper bound for the velocity of the
head, where $j_i$ is the capacity of the largest
barrier the head has already left behind.
This upper bound obviously changes at such a barrier which is larger than 
any of the barriers on its left-hand side. 
The $m$th barrier with this property, 
the so-called $m$th {\it limiting barrier} has a typical
distance from the origin $n_m\sim O(2^m)$ and a typical trapping time 
$\tau_m\sim n_m^{1/\mu}$.  
Let us consider the $m$th limiting barrier, assume that $m\gg 1$ 
and estimate the time $t$ that elapses until, starting from here, 
the head reaches the next limiting barrier, which
is located a distance of $O(n_m)$ away.
As aforesaid, the current of the limiting barrier provides a lower bound
$T_j$ for the travelling time: $t>T_j\sim \frac{n_m}{j_m(l+1)}$. 
On the other hand, the waiting time of the head at a barrier
is at least in the order of the one-particle waiting time, therefore
another lower bound is given by the single-particle travelling time: 
$t>T_1\sim \sum_i \tau_i$. Since the delay of the head at
trapping regions and the inflow of particles at the $m$th limiting
barrier take place simultaneously, we may write $T_j,T_1<t<T_j+T_1$
and conclude that the travelling time is composed of two
contributions:
\be 
t\sim O(T_j)+O(T_1).
\label{travt}
\ee         
The first term in eq. (\ref{travt}) scales with the distance $n_m$ as 
$T_j\sim n_m^{1+\frac{1}{(l+2)\mu}}$. The second term is proportional
to $n_m$ for $\mu >1$, while it is $T_1\sim n_m^{1/\mu}$ for
$\mu <1$. Comparing the two contributions, we obtain that, for 
$\mu>\mu_l^*\equiv\frac{l+1}{l+2}$, the travelling time is dominated
by $T_j$, whereas for $\mu<\mu_l^*$ it is dominated by $T_1$ for large $n_m$. 
Therefore, inverting these relations, we obtain that the head of the
string advances for $\mu>\mu_l^*$ asymptotically as 
\be 
\overline{\langle x\rangle}  \sim t^{\frac{(l+2)\mu}{(l+2)\mu+1}}, 
\label{diff1}
\ee
while if $\mu<\mu_l^*$, we have
\be 
\overline{\langle x\rangle}  \sim t^{\mu}
\label{diff2}
\ee
for large $t$.
The diffusion of the head is thus anomalous and the diffusion
exponent varies continuously with $\mu$. At $\mu=\mu_l^*$ a change
occurs in the dependence of the diffusion exponent on $\mu$ and 
below this value, the motion of the head follows the diffusion law
characteristic of the one-particle problem. 

The motion of the head between two limiting barriers is much more
complex on a microscopic length scale. 
At small trapping regions with a waiting time 
$\tau_i<j_m^{-1}\sim \tau_m^{-1/(l+2)}$, the 
head is delayed until the trapping region is filled up with particles. 
The release time of the head at large barriers with
$\tau_i>j_m^{-1}$, the number of which is
$n_m\int_{j_m^{-1}}^{\infty}\tau^{-\mu-1}d\tau \sim
n_m\tau_m^{-\mu/(l+2)} \sim n_m^{(l+1)/(l+2)}$,
is $O(\tau_i)$. During this time, particles pile up and a queue 
of length $\xi_i\sim \tau_m^{-\frac{1}{l+2}}\tau_i$ forms 
behind such a barrier. 
After the head was released, its velocity is temporarily
determined by the capacity of the barrier instead of that of the true
limiting barrier until the (the excess part of the) queue
dissolves, which takes a
time $\xi_i/j_i \sim (\frac{\tau_i}{\tau_m})^{\frac{1}{l+2}}\tau_i$.
For the length of the queue behind the $m+1$st limiting barrier, we
obtain 
$\xi_{n_{m+1}}\sim \tau_m^{(l+1)/(l+2)}\sim n_m^{\mu_l^*/\mu}$. 
Thus for $\mu>\mu_l^*$, the length of the queue piling up behind a limiting
barrier until the head is released is typically only a vanishing fraction
of the distance from the preceding limiting barrier, while for
$\mu<\mu_l^*$, the domain behind the head is almost completely filled up with 
particles. 
In fact, in the latter case, also queues forming at non-limiting but
large barriers may extend to the preceding limiting barrier. The
queue temporarily blocks the inflow of particles here, which
explains the increase of travelling time of the head compared to that
dictated solely by the current of the limiting barrier.  
We see again from the dynamics of the queues forming at trapping regions that
the properties of the system change at the value $\mu_l^*$ of the
control parameter. 

\begin{figure}[h]
\begin{center}
\includegraphics[width=0.45\linewidth]{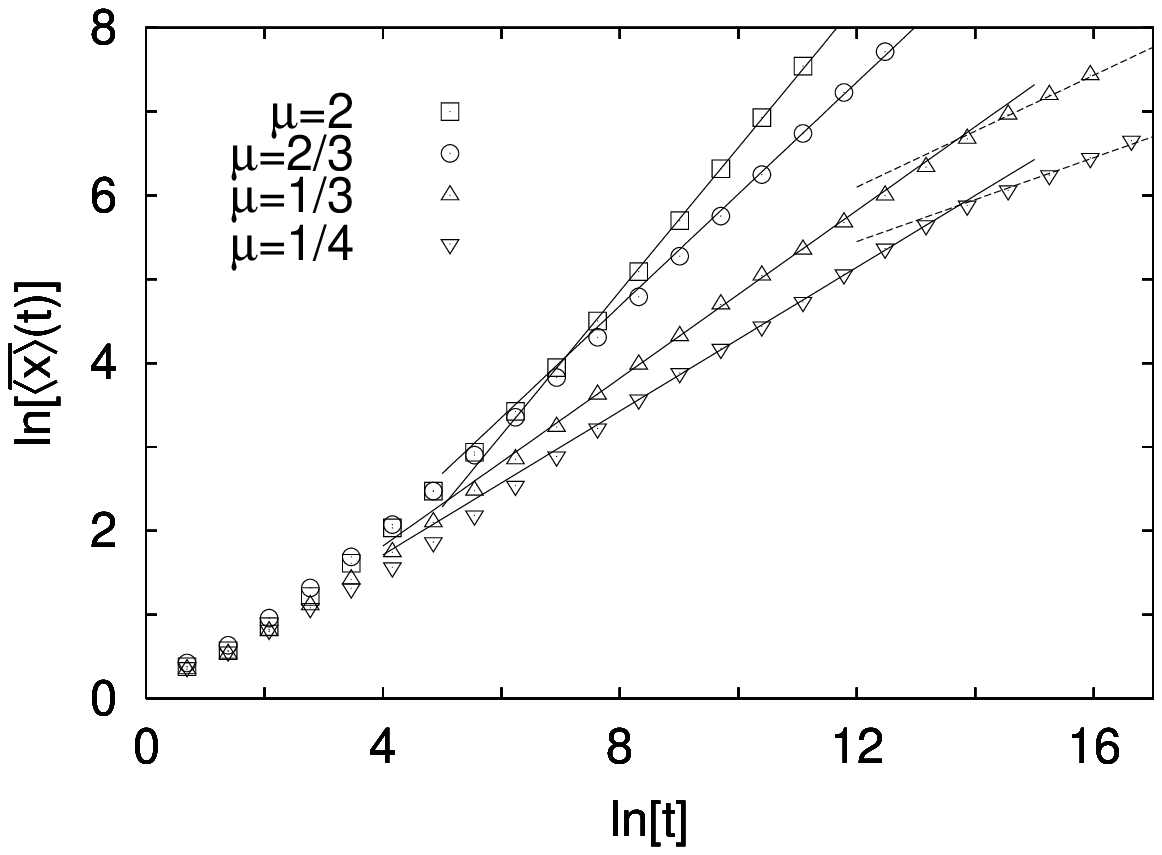}
\includegraphics[width=0.45\linewidth]{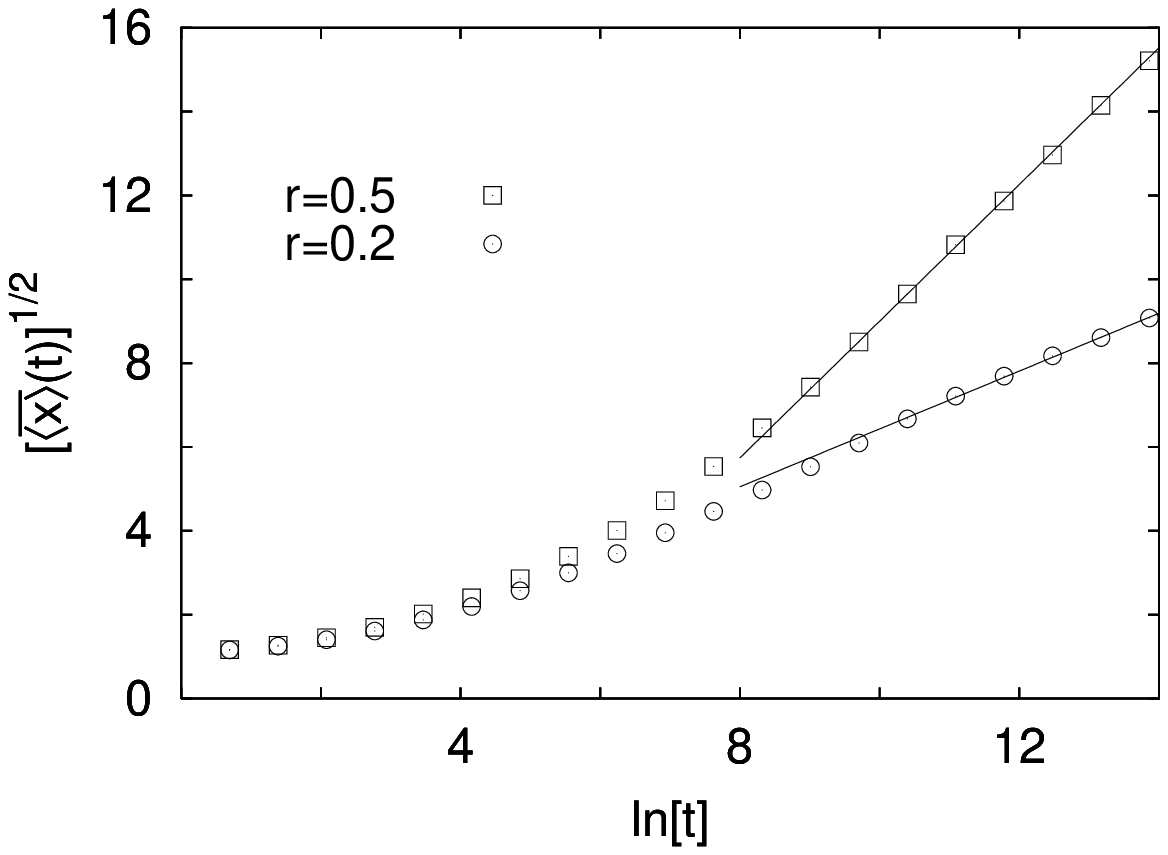}
\caption{\label{numinv} Results of numerical simulations. Left panel: 
the logarithm of the average position of the head 
plotted against the logarithm of time in the model with $l=1$ for different values of 
the control parameter. 
Binary randomness was used with parameters 
$c=0.2$, $r=0.5$; $c=0.2$, $r=1/8$; $c=1/3$, $r=1/8$; $c=1/3$,
$r=1/16$, where the control parameter takes the values
$\mu=2,2/3,1/3$ and $1/4$,
respectively. The average was performed over a few hundred independent
samples and $3-20$ runs for each sample. The slope of the solid lines
is $\frac{(l+2)\mu}{(l+2)\mu+1}$, while that of dashed lines is $\mu$.  
Right panel: Square root of the average position of the head plotted against
the logarithm of time in the model with $l=1$ in the unbiased case
$\mu =0$. 
The parameters $c=0.5$
and $r=0.5,0.2$ were used and the data were averaged over $200$
independent samples and $20$ runs per sample.} 
\end{center}
\end{figure}

We have performed numerical simulations and measured the
time-dependence of the average displacement of the head for different
values of the control parameter. Results are shown
in Fig. \ref{numinv}. As can be seen, for $\mu\ge\mu_l^*$, the
diffusion exponent is in good agreement with the predictions. 
For $\mu<\mu_l^*$ and for short times, the diffusion seems to follow 
the law found for $\mu>\mu_l^*$ (eq. (\ref{diff1})) and for longer
times a crossover can be observed to the true asymptotic behaviour
(eq. (\ref{diff2})). This means that for short times,
the second term in eq. (\ref{travt}) is suppressed by the first one, but 
ultimately, it dominates the travelling time 
since it grows with a greater exponent.    

In the limit $\mu\to 0$,
the diffusion exponent tends to zero and we expect that for $\mu =0$, 
the displacement $\overline{\langle x\rangle}(t)$ 
grows slower than any power of $t$. 
The unbiased case is out of the scope of the above theory, nevertheless, 
we can estimate the leading order behaviour of 
$\overline{\langle x\rangle}(t)$ by taking into account that    
the sojourn time $t$ of a random walker in a domain of size $\xi$ scales as 
$\ln t\sim \sqrt{\xi}$. Thus, we expect that the average displacement
grows as  
$\overline{\langle x\rangle}(t)\sim C(\ln t)^2$, where $C$ is a
non-universal constant. Results of numerical simulations are in accordance with
this relation (see Fig. \ref{numinv}).

\section{Traffic of finite open strings}
\label{finite}

In the final part of this work, we shall study finite open strings
composed of $m$ particles, which we call shortly $m$-strings.

\subsection{Steady state}

First, we are interested in the steady-state properties of
systems defined on a ring, in which the density of $m$-strings
is finite.    
This model belongs to the model class discussed in
Sec. \ref{general}, for which the steady-state results presented there
are at our disposal once the basic exponents $\mu_1$ and $\mu_2$ are known.  

\begin{figure}[h]
\begin{center}
\includegraphics[width=0.5\linewidth]{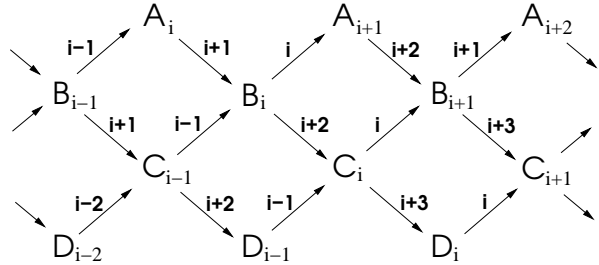}
\caption{\label{network} Network of transitions for an
  $m$-string with $m=2$ and $l=3$. An arrow with index $i$ symbolises an
  allowed transition in the direction of the arrow with rate $p_i$ and in
  the opposite direction with rate $q_{i+1}$.} 
\end{center}
\end{figure}

In the high-density phase, the active holes which are responsible for
the current in the inter-barrier regions are simple holes,
consequently, we have $\mu_1=\mu$. 
In the low-density phase, the
active particles are $m$-strings moving in the inter-barrier domains.    
Thus we need the diffusion exponent of an $m$-string characterised 
by the parameters $m$ and $l$.
This can be obtained by constructing the
network of possible transitions of such systems. 
Let us consider the simplest case, $m=2$, a ``dimer'' and denote 
the state when the left-hand side particle of the dimer is located 
at site $i$ 
and the right-hand side particle at site 
$i+1,i+2,i+3,\dots$ by $A_i,B_i,C_i,\dots$, respectively. 
The network of transitions
for $l=3$ is depicted in Fig. \ref{network} and the corresponding networks for
$l=2$ ($l=1$) are obtained by deleting the nodes $D_i$ ($D_i$ and
$C_i$) and the links connected with them.   
The conclusions drawn through this example can be
easily extended to general $m$ and $l$. 
As can be seen, the diffusion of an $m$-string can be regarded
as a random walk on a
disordered quasi-one-dimensional lattice. The walker has several
alternative paths to get from one site to another one, however, the
potential difference defined by eq. (\ref{pot}) is the same along all 
paths in an infinite system. Thus, a single-valued potential can be 
introduced here, just as for a one-dimensional chain.  
Therefore, when getting from one site to a remote
one, the trapping regions cannot be ``walked round'' and in the point of
view of large-scale properties, the existence of alternative paths is 
irrelevant.
Moreover, one can easily make sure that following a path in the
network between two distant points, each type of link is contained
$m$-times in the path. See, for example, the path
$A_i\to B_i\to A_{i+1}\to B_{i+1}\to\dots$ in the figure, in which the
links of the original lattice are contained doubly. 
Consequently, the effective height of a barrier felt by an $m$-string
is $m$ times greater than the true height. 
This can also be seen in a more heuristic way by taking into account
that one step of an $m$-string amounts to $m$ steps  
of its constituent particles.       
Thus, we conclude that $f_2=m$ and the exponent governing the
diffusion of an $m$-string is $l$-independent and is given by $\mu_2=\mu/m$.     
The results in eq. (\ref{Jgen})-(\ref{agen}) are thus valid for this model    
with $\mu_1=\mu$ and $\mu_2=\mu/m$. For example, for the finite-size
scaling of the current we obtain:
\be 
J_L\sim L^{-\frac{1}{\mu}\frac{m}{m+1}}.
\label{Jstring}
\ee

\subsection{Invasion}

In the following, a problem similar to that discussed in
Sec. \ref{diffusion} will be examined, 
i.e. we consider a semi-infinite lattice, which is initially empty,
and at the first lattice site $m$-strings enter (``invade'') the
lattice with a finite entrance rate 
and we are interested in the advance of the $m$-string first entered the
system, as well as the evolution of the number of particles on the
lattice.  
Although, the steady-state behaviour could be handled together with the
problem of a closed string within a common formalism, 
the problem of invasion of $m$-strings must be treated separately
from the diffusion of a semi-infinite string  
and it has much in common with the invasion of uncoupled
particles ($m=1$), which has been studied in Ref. \cite{jsi}. 

From now on, we shall call the $m$-strings particles.  
First, it is clear that the first particle advances with a constant
velocity if $\mu_2>1$ ($\mu >m$). 
Let us assume that $0<\mu_2<1$ and consider the $k$th limiting
barrier far from the entrance site, i.e. $k\gg 1$, and estimate the
travelling time of the first particle between the $k$th and the $k+1$st
limiting barrier. 
In this domain, the first particle is caught up by the following
particles only at sufficiently large barriers, for which 
$\tilde\tau_2>j_k^{-1}$, where $j_k\sim n_k^{-1/(\mu_1+\mu_2)}$ 
is the current supplied by the $k$th limiting barrier.
The number of such barriers is $n_B\sim  n_k^{\mu_1/(\mu_1+\mu_2)}$.
Such a  trapping region is filled up with particles in a period of 
$O(j_k^{-1})$  to a level at which the escape time of the first
particle is in the same order of magnitude, $O(j_k^{-1})$.  
The waiting time of the first particle at such a barrier 
is thus $O(n_k^{1/(\mu_1+\mu_2)})$. 
The travelling time of the first particle between two such adjacent barriers,
which are located a distance  
$\xi_0 \sim n_k/n_B\sim n_k^{\mu_2/(\mu_1+\mu_2)}$ apart is 
in the same order of magnitude: 
$t_0\sim \xi_0^{1/\mu_2}\sim n_m^{1/(\mu_1+\mu_2)}$.
Thus the total travelling time between the $k$th and the $k+1$st limiting
barrier is 
$t\sim n_Bn_k^{1/(\mu_1+\mu_2)}\sim n_k^{(1+\mu_1)/(\mu_1+\mu_2)}$. 
Inverting this relation, we obtain that 
the advance of the first particle follows the asymptotic law:
\be 
\overline{\langle x\rangle}  \sim t^{\frac{1+1/m}{1+1/\mu}},   \qquad \qquad 
\mu<m.
\ee 

Nevertheless, the average number of particles $N(t)$ can be shown
to grow slower. The number of particles present in the
system is approximately equal to the length of the high-density segment, which
extends from the entrance site to an advancing front \cite{jsi}.   
When the front is located at a limiting barrier in a distance $N$ 
from the entrance, the inflow of particles is controlled by the
capacity of this barrier. For the growth rate of $N$, we may thus write 
$\frac{dN}{dt}\sim N^{-1/(\mu_1+\mu_2)}$, which yields
\be
N(t)\sim t^{\frac{1+1/m}{1+1/m+1/\mu}}.
\ee
So, the total number of particles in the system grows with a
smaller exponent than the displacement of the first particle. 
Such a ``dispersion'' is
obviously not possible in the case of a semi-infinite
string since particles are coupled. 
Setting $m=1$ in the above expressions we recover known results for
the ASEP.     

In the unbiased case ($\mu=0$), the advance of the leading particle is
expected to be ultra-slow and to follow the logarithmic scaling law
characteristic of Sinai diffusion.
 
\section{Discussion}  
\label{discussion}

As it was outlined in Ref. \cite{jsi}, the holes in the 
closed string with $l=1$ and with hop rates $p_i,q_i$ can be regarded 
as large particles of size $d=2$  with hop rates $\tilde p_i=q_{i+1}$,
$\tilde q_i=p_{i-1}$.
Although, for $l>1$, a strict mapping between the two models does not
exist, the phenomenological results obtained in this work show that
the basic exponents $\mu_1,\mu_2$ of the closed string with parameter $l$  
are related to the basic exponents $\tilde\mu_1,\tilde\mu_2$ of an 
exclusion process with particles of size $d=l+1$ in the following way:
$\mu_1=\tilde\mu_2$, $\mu_2=\tilde\mu_1$.   

Comparing the exponent governing the decay of the steady-state current
of a closed string to that of the ASEP, we see that the former is
smaller for any $\mu>0$, i.e. the coupling between particles
facilitates the transport in this set-up. Moreover, the exponent decreases
monotonously with $l$ and finally tends to zero in the limit $l\to
\infty$\footnote{We remind the reader that our results are valid in 
scaling regime $\ln L\gg l$.}.
When not all particles are coupled with each other but
they form finite $m$-strings, the coupling has the opposite effect:
With increasing $m$, the transport slows down more and more 
compared to the ASEP both in the steady state
and in the case of the invasion.     

Contrary to the problem of invasion of uncoupled particles, 
in the case of a semi-infinite string, both the displacement of the
head and the number of particles in the system grows obviously with
the same power of time.  This exponent is larger than
that describing the increase of the number of particles in the ASEP 
if $\mu>1/2$, meaning that the coupling is 
favourable in this case in the point of view of the bulk of
particles. 
However, if $\mu<1/2$, the inflow of particles is slowed down 
by the coupling.  
As far as the first particle is concerned, it diffuses 
faster in the ASEP if $\mu>\mu_l^*$, otherwise the diffusion exponents
are equal in the two models.   
The influence of the parameter $l$ on the transport shows a
tendency similar to that in the closed string:
The diffusion exponent increases monotonously with $l$
(except of the regime  $\mu<\mu_l^*$, where it is $l$-independent) 
and tends to $1$ if $l\to\infty$. 

The diffusion of the semi-infinite string can be speeded up 
by ``pulling'' the head of the string, that means when 
the first particle is not allowed to hop to the left. 
In this case, one can show that the time scale of overcoming a
barrier is reduced from $O(\tau)$ to $O(\tau^{1/(l+2)})$, which
results in that the diffusion law in eq. (\ref{diff1}) 
is valid in the entire biased phase $\mu >0$. 
Comparing this model to the invasion of uncoupled particles,  
the diffusion of the pulled string is faster than that of
the first particle in case of the ASEP only if
$\mu<\frac{l}{l+2}$, while the inflow of particles is faster for the pulled
string for any $\mu$. 
       
Relaxing the hard-core exclusion condition in the ASEP, i.e. 
prescribing for the positions of consecutive particles the weaker 
condition $x_i\le x_{i+1}$ instead of $x_i<x_{i+1}$, we arrive at a 
zero-range process.
Doing so for the model with coupled particles but keeping the attractive
interaction $x_{i+1}-x_{i}\le l+1$, we obtain a model which is no
longer a ZRP. Due to the absence of hard-core repulsion, 
the high-density phase in the steady state of the closed string 
shrinks to the minimum in the largest trapping region,
where a condensate forms and the low-density phase extends to the whole
system. The current is thus expected to vanish as 
$J_L\sim L^{-1/\mu_2}\sim  L^{-\frac{1}{(l+1)\mu}}$.   

Finally, we mention that, as a possible generalisation of the model 
with limited inter-particle spacings, 
one could apply elastic pair-interactions by the help of which 
the crossover between the model with uncoupled particles 
(ASEP) and that studied in the present work could be investigated.

\ack
This work has been supported by the  
National Office of Research and Technology under Grant No. ASEP1111.


\end{document}